# Taming the divergent terms that occur during adiabatic switching in perturbation theory.

Author: Dan Solomon
Address: University of Illinois at Chicago, Physics Department, 845 W. Taylor Street, Chicago IL, 60607.
Email: dsolom2@uic.edu
Date: August 27, 2016

**Abstract.**
A potential problem with adiabatic switching in perturbation theory is that divergent terms appear in the series solution. An example of this was presented by C. Brouder et al [4] for a simple 2 state system where the evolution of system in the presence of a time dependent perturbation was considered. One of their results is that the evolution operator has no well-defined limit for adiabatic switching. We will rework this problem to show that for adiabatic switching the evolved states are well-defined with any divergences being absorbed in a time independent phase factor which can be removed. These results will then be applied to the more general problem of a system with an arbitrary number of states. It will be shown that for this case, also, the potentially divergent terms all appear in a time-independent phase factor.

## 1. Introduction.

The adiabatic theorem provides a way for the eigenstates of the Hamiltonian $H = H_0 + H_I$ to be derived from the eigenstates of $H_0$ [1][2]. To illustrate this concept first give the interaction term a time dependence by replacing $H_1$ with $e^{\varepsilon t} H_1$ where $\varepsilon > 0$ to obtain $H(t) = H_0 + e^{\varepsilon t} H_1$. At $t \to -\infty$ the Hamiltonian will be $H(-\infty) = H_0$. Assume at this initial time the system is in the state $\Upsilon_a$ which is an eigenstate of $H_0$. The system will evolve in time according to the Schrödinger equation,

$$i \frac{\partial}{\partial t} \Psi(t) = H(t) \Psi(t) \qquad (1.1)$$

Let the solution to this be given by $\psi_{a,\varepsilon}(t)$ where the initial condition is $\psi_{a,\varepsilon}(-\infty) = \Upsilon_a$. At $t = 0$ the wave function will have evolved into the state $\psi_{a,\varepsilon}(0)$. Now at $t = 0$ the Hamiltonian is given by $H(0) = H_0 + H_1$. What is the relationship, if any, between the state $\psi_{a,\varepsilon}(0)$ and the eigenstates of $H(0)$? According to the adiabatic theorem in the limit $\varepsilon \to 0$ the state $\psi_{a,\varepsilon}(0)$ is an eigenstate of $H(0)$.

However there is a potential problem. If the state $\psi_{a,\varepsilon}(0)$ is derived using perturbation theory then there is the possibility that in the limit $\varepsilon \to 0$ the solution will become highly divergent [3]. This problem was examined by Brouder et al [4] who consider an exactly solvable 2 state model and claim to show that $\psi_{a,\varepsilon}(0)$ has no limit when $\varepsilon \to 0$. We will rework the problem and show that in the limit $\varepsilon \to 0$ a perfectly reasonable solution exists with all the divergent terms being absorbed in a time independent phase factor. Furthermore the adiabatic theorem is confirmed for this case because, in the limit $\varepsilon \to 0$, $\psi_{a,\varepsilon}(0)$ will be an eigenfunction of $H(0) = H_0 + H_1$. We will also show by direct calculation that, for the general case of an arbitrary number of states, the divergent terms that occur during adiabatic switching also only appear in a time-independent phase factor.

## 2. The model system.

C. Brouder et al [4] considered a two state system whose Hamiltonian is $H = H_0 + xV$ where,

$$H_0 = \begin{pmatrix} \mu - \delta & 0 \\ 0 & \mu + \delta \end{pmatrix}, \quad V = \begin{pmatrix} 0 & 1 \\ 1 & 0 \end{pmatrix}, \tag{2.1}$$

and where $\delta > 0$ and $x > 0$. The eigenstates are solutions to $(H_0 + xV)\Psi_n = E_n \Psi_n$ where $n = 0$ or $1$ and $E_n$ is the eigenvalue associated with $\Psi_n$. Using (2.1) in this we obtain,

$$\begin{pmatrix} \mu - \delta & x \\ x & \mu + \delta \end{pmatrix} \begin{pmatrix} r_1 \\ r_2 \end{pmatrix} = E \begin{pmatrix} r_1 \\ r_2 \end{pmatrix} \tag{2.2}$$

The $r_1$ and $r_2$ are constants and satisfy the normalization condition, $|r_1|^2 + |r_2|^2 = 1$. The two eigenstate solutions are,

$$\Psi_0(x) = N(x) \begin{pmatrix} 1 \\ \Delta E(x)/x \end{pmatrix}, \quad E_0(x) = (\mu - \delta) - |\Delta E(x)| \tag{2.3}$$

and,

$$\Psi_1(x) = N(x) \begin{pmatrix} -\Delta E(x)/x \\ 1 \end{pmatrix}, \quad E_1(x) = (\mu + \delta) + |\Delta E(x)| \tag{2.4}$$

where $\Delta E(x) = \delta - \sqrt{\delta^2 + x^2}$ and $N(x) = 1 \big/ \left( \sqrt{1 + (\Delta E(x)/x)^2} \right)$. If $x = 0$ the unperturbed states are,

$$\Upsilon_0 = \begin{pmatrix} 1 \\ 0 \end{pmatrix}, \quad E_0(0) = \mu - \delta; \quad \Upsilon_1 = \begin{pmatrix} 0 \\ 1 \end{pmatrix}, \quad E_1(0) = \mu + \delta \tag{2.5}$$

Of the two solutions the one with the lowest energy is $\Psi_0(x)$. Therefore this will be considered the "vacuum" state and $\Psi_1(x)$ will be considered the excited state. For the unperturbed system $(x = 0)$ the vacuum state is $\Upsilon_0$. For latter convenience we will rewrite (2.3) as,

$$\Psi_0(x) = N(x)\left[ \Upsilon_0 + (\Delta E(x)/x)\Upsilon_1 \right] \tag{2.6}$$

## 3. Time dependent perturbation.

Consider our 2-state model system evolving in the presence of the time dependent Hamiltonian,

$$H(t) = H_0 + xe^{\varepsilon t}V \tag{3.1}$$

where $\varepsilon > 0$. Let $\psi_0(t;\varepsilon)$ be the wave function of this system where the initial state at $t \to -\infty$ is $\Upsilon_0$. The wave function, $\psi_0(t;\varepsilon)$, evolves in time according to the Schrödinger Eq. (1.1). We will solve this problem using the results of Ref. [4] and consider how the solution behaves as $\varepsilon \to 0$.

As discussed in [4] the solution to the Schrödinger equation is given by,

$$\psi_0(t;\varepsilon) = e^{-iH_0 t} U(t) \Upsilon_0 \tag{3.2}$$

where $U(t)$ is the evolution operator in the interaction picture. We denote,

$$U(t) = \begin{pmatrix} a(t) & b(t) \\ c(t) & d(t) \end{pmatrix} \tag{3.3}$$

where,

$$ia' = xe^{-i(2\delta + i\varepsilon)t}c, \quad ic' = xe^{i(2\delta - i\varepsilon)t}a, \quad ib' = xe^{-i(2\delta + i\varepsilon)t}d, \quad id' = xe^{i(2\delta - i\varepsilon)t}b \tag{3.4}$$

and where $a' = da(t)/dt$ and the initial conditions are $a(-\infty) = d(-\infty) = 1$ and $b(-\infty) = c(-\infty) = 0$. Use (3.3) in (3.2) to obtain,

$$\psi_0(t;\varepsilon) = e^{-iH_0 t}U(t)\Upsilon_0 = e^{-iH_0 t}\begin{pmatrix} a(t) \\ c(t) \end{pmatrix} = e^{-i\mu t}\begin{pmatrix} e^{+i\delta t}a(t) \\ e^{-i\delta t}c(t) \end{pmatrix} = e^{-i(\mu-\delta)t}\left(a(t)\Upsilon_0 + \frac{ia'(t)}{x}e^{-\varepsilon t}\Upsilon_1\right) \tag{3.5}$$

From the above it can be shown that

$$a'' + (2i\delta - \varepsilon)a' + x^2 e^{2\varepsilon t}a = 0, \tag{3.6}$$

The solution to this is given by [4] as,

$$a(t) = C_2\left(\frac{x}{\varepsilon}\right)e^{\varepsilon vt} J_{-v}\left(\frac{xe^{\varepsilon t}}{\varepsilon}\right), \tag{3.7}$$

where $C_2 = 2^{-v}\Gamma(1-v)$. According to [4] the series expansion of the Bessel function yields,

$$a(t) = 1 + \sum_{k=1}^{\infty} \frac{(-s^2/4)^k}{k!\prod_{j=1}^{k}(j-v)}, \tag{3.8}$$

where $s = xe^{\varepsilon t}/\varepsilon$ and $v = 1/2 - i\delta/\varepsilon$. The problem with this expansion is that it becomes highly divergent as $\varepsilon \to 0$. This can be easily seen by writing out the first few terms to obtain,

$$a(t) = \left[1 - \frac{x^2 e^{2\varepsilon t}/\varepsilon}{2(\varepsilon + 2i\delta)} + \frac{x^4 e^{4\varepsilon t}/\varepsilon^2}{8(\varepsilon + 2i\delta)(3\varepsilon + 2i\delta)} - \frac{x^6 e^{6\varepsilon t}/\varepsilon^3}{48(\varepsilon + 2i\delta)(3\varepsilon + 2i\delta)(5\varepsilon + 2i\delta)} + \ldots\right] \tag{3.9}$$

As can be seen the nth term in the expansion will have a $1/\varepsilon^n$ factor which will be highly divergent in the limit $\varepsilon \to 0$. Therefore in the limit $\varepsilon \to 0$ it would seem that $a(t)$ is undefined. However it will be shown in the next section that this problem can be reworked so that all the divergent terms are located in a time-independent phase factor which may be discarded to obtain a perfectly well defined solution.

## 4. An Alternative Calculation.

We will rework the problem presented in the last section and show that $a(t)$ can be evaluated so that it makes sense in the adiabatic limit, $\varepsilon \to 0$. The divergent terms will be all contained in a time-independent phase factor that can be readily removed to obtain a well-defined expression. Express $a(t)$ in terms of a new function $f(t)$ as,

$$a(t) = \exp\left(-i\frac{f(t,\varepsilon)}{\varepsilon}\right) \tag{4.1}$$

where,

$$f(t,\varepsilon) = \varepsilon \int_{-\infty}^{t} g(t_1,\varepsilon) dt_1 \tag{4.2}$$

We can solve for $g(t,\varepsilon)$ by using the above relationships in (3.6) to obtain,

$$g' - ig^2 + (2i\delta - \varepsilon)g = -ix^2 e^{2\varepsilon t} \tag{4.3}$$

Next define $\lambda(t) = xe^{\varepsilon t}$ and $\tilde{g}(\lambda(t)) = g(t,\varepsilon)$ and use $d\tilde{g}/dt = \varepsilon\lambda \, d\tilde{g}/d\lambda$ in the above to obtain,

$$\varepsilon\lambda\frac{d\tilde{g}}{d\lambda} - i\tilde{g}^2 + (2i\delta - \varepsilon)\tilde{g} = -i\lambda^2 \tag{4.4}$$

Expand $\tilde{g}$ in terms of $\lambda^2$ to obtain,

$$\tilde{g}(\lambda) = \sum_{n=1}^{\infty} \lambda^{2n} \tilde{g}_n(\varepsilon); \quad g(t,\varepsilon) = \sum_{n=1}^{\infty} x^{2n} e^{2n\varepsilon t} \tilde{g}_n(\varepsilon) \tag{4.5}$$

where $\tilde{g}_n(\varepsilon)$ are time-independent expansion coefficients. Use this in (4.4) to obtain,

$$\sum_{n=1}^{\infty} \lambda^{2n} \tilde{g}_n \left(2i\delta + (2n-1)\varepsilon\right) = i\sum_{n=2}^{\infty} \lambda^{2n} \sum_{m=1}^{n-1} \tilde{g}_{n-m}\tilde{g}_m - i\lambda^2 \tag{4.6}$$

From this we obtain,

$$\tilde{g}_1(\varepsilon) = \frac{-i}{(2i\delta + \varepsilon)}; \quad \tilde{g}_n(\varepsilon) = \frac{i\sum_{m=1}^{n-1} \tilde{g}_{n-m}(\varepsilon)\tilde{g}_m(\varepsilon)}{(2i\delta + (2n-1)\varepsilon)} \text{ for } n = 2,3,\ldots,\infty \tag{4.7}$$

The first few of these terms are given in Appendix 1. Use (4.5) in (4.2) and perform the integration to obtain,

$$f(t;\varepsilon) = \sum_{n=1}^{\infty} \frac{\lambda^{2n}}{2n} \tilde{g}_n(\varepsilon) = \sum_{n=1}^{\infty} \frac{x^{2n} e^{2n\varepsilon t}}{2n} \tilde{g}_n(\varepsilon) \tag{4.8}$$

where we have used $\int_{-\infty}^{t} \lambda^{2n} dt_1 = \left(\lambda^{2n}/2n\varepsilon\right)$.

We are interested in evaluating $a(t)$ in the situation where $\varepsilon \to 0$. Note that in this limit the $\tilde{g}_n(\varepsilon)$ are well defined. For example, consider $\tilde{g}_2(\varepsilon)$ which is given in Appendix 1. In the limit $\varepsilon \to 0$, $\tilde{g}_2(\varepsilon)$ becomes $\tilde{g}_2(0) = -i/(2i\delta)^3$. Therefore the only possible divergent terms are due to the $(1/\varepsilon)$ term that multiplies $f(t)$. For sufficiently small $\varepsilon/\delta$ we can expand $\tilde{g}_n(\varepsilon)$ as a power series in $\varepsilon$ to obtain,

$$\tilde{g}_n(\varepsilon) = \tilde{g}_n(0) + r_n(\delta)\varepsilon + \sum_{k=2}^{\infty} \frac{\varepsilon^k}{k!}\left(\frac{d^k \tilde{g}_n(\varepsilon)}{d\varepsilon^k}\right)_{\varepsilon=0} \qquad (4.9)$$

where,

$$r_n(\delta) = \left(\frac{d\tilde{g}_n(\varepsilon)}{d\varepsilon}\right)_{\varepsilon=0} \qquad (4.10)$$

Therefore,

$$\frac{f(t,\varepsilon)}{\varepsilon} = \sum_{n=1}^{\infty} \frac{x^{2n} e^{2n\varepsilon t}}{2n}\left(\frac{\tilde{g}_n(0)}{\varepsilon} + r_n(\delta) + \sum_{k=2}^{\infty} \frac{\varepsilon^{k-1}}{k!}\left(\frac{d^k \tilde{g}_n(\varepsilon)}{d\varepsilon^k}\right)_{\varepsilon=0}\right) \qquad (4.11)$$

We can write this as,

$$\frac{f(t,\varepsilon)}{\varepsilon} = \frac{F_a(t,\varepsilon)}{\varepsilon} + iF_b(t,\varepsilon) + F_c(t,\varepsilon) \qquad (4.12)$$

where,

$$F_a(t,\varepsilon) = \sum_{n=1}^{\infty} \frac{x^{2n} e^{2n\varepsilon t} \tilde{g}_n(0)}{2n}, \quad F_b(t,\varepsilon) = -i\sum_{n=1}^{\infty} \frac{x^{2n} e^{2n\varepsilon t} r_n(\delta)}{2n}, \quad F_c(t,\varepsilon) = \sum_{n=1}^{\infty} \frac{x^{2n} e^{2n\varepsilon t}}{2n}\sum_{k=2}^{\infty} \frac{\varepsilon^{k-1}}{k!}\left(\frac{d^k \tilde{g}_n(\varepsilon)}{d\varepsilon^k}\right)_{\varepsilon=0}$$

(4.13)

This allows us to write,

$$a(t) = \exp\left(-i\frac{F_a(t,\varepsilon)}{\varepsilon}\right)\exp(F_b(t,\varepsilon))\exp(-iF_c(t,\varepsilon)) \qquad (4.14)$$

In the limit that $\varepsilon \to 0$ we have that,

$$\frac{F_a(t,\varepsilon)}{\varepsilon} \underset{\varepsilon \to 0}{=} \sum_{n=1}^{\infty}\left(\frac{x^{2n} \tilde{g}_n(0)}{2n\varepsilon} + x^{2n}\tilde{g}_n(0)t\right), \quad F_b(t,\varepsilon) \underset{\varepsilon \to 0}{=} F_b(0,0) = -i\sum_{n=1}^{\infty} \frac{x^{2n} r_n(\delta)}{2n}, \quad F_c(t,\varepsilon) \underset{\varepsilon \to 0}{=} 0$$

(4.15)

where we have used $e^{2n\varepsilon t} \underset{\varepsilon \to 0}{=} (1 + 2n\varepsilon t + O(\varepsilon^2))$. Define $\Delta E_a$ by,

$$\Delta E_a = \sum_{n=1}^{\infty} x^{2n} \tilde{g}_n(0) \qquad (4.16)$$

Therefore in the limit $\varepsilon \to 0$,

$$a(t) \underset{\varepsilon \to 0}{=} \exp\left(-i\frac{F_a(0,0)}{\varepsilon}\right)\exp(-i\Delta E_a t)\exp(F_b(0,0)) \qquad (4.17)$$

In the above $F_a(0,0)$, $\Delta E_a$, and $F_c(0,0)$ are all real and time-independent. There the only divergent term is due to a time-independent phase factor.

Now we can use all this in Eq. (3.5) to obtain the wave function $\psi_0(t;\varepsilon)$ in the limit $\varepsilon \to 0$ as,

$$\psi_0(t;\varepsilon) \underset{\varepsilon \to 0}{=} e^{-i(\mu-\delta)t} \exp\left(-i\frac{F_a(0,0)}{\varepsilon}\right) \exp(-i\Delta E_a t) \exp(F_b(0,0)) \left(\Upsilon_0 + \frac{\Delta E_a}{x}\Upsilon_1\right) \tag{4.18}$$

Next determine $\Delta E_a$ which is given in Eq. (4.16). First refer to (4.7) to obtain,

$$\tilde{g}_1(0) = \frac{-1}{2\delta}; \quad \tilde{g}_n(0) = \frac{\sum_{m=1}^{n-1} \tilde{g}_{n-m}(0)\tilde{g}_m(0)}{2\delta} \quad \text{for } n = 2, 3, \ldots, \infty \tag{4.19}$$

Using this result along with (4.16) it can be readily shown that $\Delta E_a$ satisfies the equation,

$$\Delta E_a^2 - 2\delta \Delta E_a - x^2 = 0 \tag{4.20}$$

We can solve this directly to obtain,

$$\Delta E_a = \delta - \sqrt{\delta^2 + x^2} \tag{4.21}$$

This is the same as $\Delta E(x)$ defined at the end of Section 2. Therefore in the following $\Delta E_a$ will be written without the subscript as $\Delta E(x)$.

Next determine $F_b(0,0)$. Since unitarity has been preserved throughout and $F_b(0,0)$ is real we can use $\psi_{S,vac}(t)^\dagger \psi_{S,vac}(t) = 1$ to expect that,

$$e^{F_b(0,0)} = 1 \Big/ \sqrt{\left(1 + (\Delta E(x)/x)^2\right)} \tag{4.22}$$

This result is verified in Appendix 2. Thus (4.18) becomes,

$$\psi_0(t, \varepsilon \to 0) = \exp\left(-i\frac{F_a(0,0)}{\varepsilon}\right) e^{-i(\mu-\delta)t} \exp(-i\Delta E_a t) \Psi_0(x) \tag{4.23}$$

where $\Psi_0(x)$ is the time-independent "vacuum" state defined by Eq. (2.6). The divergent part of this expression is completely contained in a time-independent phase factor. This phase factor can be dropped so that at time $t = 0$ we obtain $\psi_0(0, \varepsilon \to 0) = \Psi_0(x)$. Therefore we have shown that the in the case of adiabatic switching the solution to the Schrodinger equation equals an eigenfunction of the time independent Hamiltonian $H(0) = H_0 + xV$.

## 5. The General Case

So far we have examined a quantum system with two energy eigenstates. In this section we will consider the general case with an arbitrary number of eigenstates. Let $|n\rangle$ be the normalized eigenvectors of the unperturbed Hamiltonian $H_0$ with corresponding eigenvalues $E_n$ where $n = 0, 1, 2 \ldots \infty$. Therefore,

$$H_0|n\rangle = E_n|n\rangle, \quad \langle n|n\rangle = 1, \quad \sum_n |n\rangle\langle n| = 1 \tag{5.1}$$

Let the full Hamiltonian be given by $H(t) = H_0 + \lambda(t)V$ where $V$ is time independent and, as before, $\lambda(t) = xe^{\varepsilon t}$ with $x > 0$ and $\varepsilon > 0$. Assume at time $t \to -\infty$ the system is in the initial state $|0\rangle$ where $|0\rangle$ is non-degenerate which means that $E_n \neq E_0$ for $n \neq 0$. Under the action of the perturbation the system evolves in time into the state $|\psi(t)\rangle$. Recall the Schrodinger equation is,

$$i\frac{\partial |\psi(t)\rangle}{\partial t} = (H_0 + \lambda(t)V)|\psi(t)\rangle; \quad \lambda(t) = xe^{\varepsilon t} \tag{5.2}$$

We want to solve the Schrodinger equation subject to the initial condition $|\psi(-\infty)\rangle = |0\rangle$. One possibility is the Dyson series. It can be shown in standard reference books [5,6] that $|\psi(t)\rangle$ is given by,

$$|\psi(t)\rangle = e^{-iH_0 t}U(t)|0\rangle; \quad U(t) = T\exp\left(-i\int_{-\infty}^{t}\lambda(t)V_I(t)\right) \tag{5.3}$$

where $T$ is the time ordering operator and $V_I(t) = \exp(iH_0 t)V\exp(-iH_0 t)$ is the perturbation in the interaction picture. Write out the Dyson series to the second order to obtain,

$$|\psi(t)\rangle = e^{-iH_0 t}\left\{1 + (-ix)\int_{-\infty}^{t}\hat{V}_I(t_1)e^{i\varepsilon t_1}dt_1 + (-ix)^2\int_{-\infty}^{t}\hat{V}_I(t_1)e^{i\varepsilon t_1}dt_1\int_{-\infty}^{t_1}\hat{V}_I(t_2)e^{i\varepsilon t_2}dt_2 + O(x^3)\right\}|0\rangle \tag{5.4}$$

Perform the integrations over time to obtain,

$$|\psi(t)\rangle = e^{-iE_0 t}\left\{1 + (-ix)\sum_n\frac{|n\rangle V_{n0}e^{\varepsilon t}}{[i(E_n - E_0) + \varepsilon]} + (-ix)^2\sum_{n,m}\frac{|n\rangle V_{nm}V_{n0}e^{2\varepsilon t}}{[i(E_n - E_0) + 2\varepsilon][i(E_m - E_0) + \varepsilon]} + O(x^3)\right\} \tag{5.5}$$

where $V_{nm} = \langle n|V|m\rangle$ and we have used $\sum_n |n\rangle\langle n| = 1$. This becomes,

$$|\psi(t)\rangle = e^{-iE_0 t}\left\{\begin{array}{l}|0\rangle + (-ix)e^{\varepsilon t}\left[\dfrac{|0\rangle V_{00}}{\varepsilon} + \sum_{n\neq 0}\dfrac{|n\rangle V_{n0}}{[i(E_n - E_0) + \varepsilon]}\right] \\ + (-ix)^2 e^{2\varepsilon t}\begin{array}{l}\left[\dfrac{|0\rangle V_{00}V_{00}}{2\varepsilon^2} + \sum_{n\neq 0}\dfrac{|n\rangle V_{n0}V_{00}}{[i(E_n - E_0) + 2\varepsilon]\varepsilon} + \sum_{m\neq 0}\dfrac{|0\rangle V_{0m}V_{m0}}{2\varepsilon[i(E_m - E_0) + \varepsilon]}\right] \\ + \sum_{n\neq 0, m\neq 0}\dfrac{|n\rangle V_{nm}V_{m0}}{[i(E_n - E_0) + 2\varepsilon][i(E_m - E_0) + \varepsilon]}\end{array} + O(x^3)\end{array}\right\} \tag{5.6}$$

When we take the limit $\varepsilon \to 0$ we see that we run into a potential problem. In (5.6) there are terms that go as $\varepsilon^{-1}$ and $\varepsilon^{-2}$ and we have only taken the Dyson series to the second order. As we take additional terms the divergences will get worse. It has been shown that this problem can be resolved by using diagrammatic techniques (see Ref. [6,7]) which show that the evolution operator $U(t)$ can be expressed as $U(t) = U_L(t)\exp[U_C(t)]$ where $U_L(t)$ is the contribution due to linked diagrams and and $U_C(t)$ is the sum of all unlinked connected diagrams and is a complex number. It has been shown by [7] that the unlinked part $U_C(t)$ can be expanded as a Laurent series in $\varepsilon$ as,

$$U_C(t) = \frac{a}{\varepsilon} + b + O(\varepsilon) \tag{5.7}$$

where $a$ is pure imaginary and $O(\varepsilon)$ are terms to the first order or higher in $\varepsilon$. Therefore all the divergence as $\varepsilon \to 0$ is contained in an imaginary phase factor.

The problem with the Dyson series is that it contains highly divergent terms which appear throughout the series and get worse as additional terms are taken. To tame these divergences additional steps are required to show that the divergences are actually all contained in a phase factor. In the following we will show that we can achieve this result by direct calculation without using diagrammatic techniques. That is, we set up the problem so that the fact that the divergences are all in a phase factor is readily apparent and no additional steps are needed. Rewrite $|\psi(t)\rangle$ as,

$$|\psi(t)\rangle = e^{-iE_0 t} e^{-iG(t)} \left( |0\rangle + |\delta\psi(t)\rangle \right); \quad G(t) = \int_{-\infty}^{t} \xi(t') dt' \tag{5.8}$$

where $\langle 0 | \delta\psi(t) \rangle = 0$. We express the wave function in this form based on the results from our original two state problem. For the two state problem we found a solution where all the divergences were confined to a phase factor. Here we our attempting to a similar result for the general problem. We hope that the divergences of the problem are confined to $G(t)$. It will be shown that this is, indeed, the case.

We must solve for $|\delta\psi(t)\rangle$ and $\xi(t)$. Use (5.8) in (5.2) to obtain,

$$\left( E_0 + \xi(t) \right)\left( |0\rangle + |\delta\psi(t)\rangle \right) + i\frac{\partial |\delta\psi(t)\rangle}{\partial t} = \left( H_0 + \lambda(t)V \right)\left( |0\rangle + |\delta\psi(t)\rangle \right) \tag{5.9}$$

Multiply by $\langle 0 |$ and use $\langle 0 | \delta\psi(t) \rangle = 0$ and $\langle 0 | H_0 = \langle 0 | E_0$ to obtain,

$$\xi(t) = \lambda(t)\left( \langle 0|V|0\rangle + \langle 0|V|\delta\psi(t)\rangle \right) \tag{5.10}$$

Next multiply (5.9) by the projection operator $Q = 1 - |0\rangle\langle 0|$ and use $Q|\delta\psi\rangle = |\delta\psi\rangle$, $Q|0\rangle = 0$, and $[H_0, Q] = 0$ to obtain,

$$\xi(t)|\delta\psi(t)\rangle + i\frac{\partial |\delta\psi(t)\rangle}{\partial t} = \left( H_0 - E_0 \right)|\delta\psi(t)\rangle + \lambda(t)QV\left( |0\rangle + |\delta\psi(t)\rangle \right) \tag{5.11}$$

Expand $\xi(t)$ and $|\delta\psi(t)\rangle$ in terms of $\lambda(t)$ according to,

$$\xi(t) = \sum_{n=1}^{\infty} \lambda(t)^n \xi_n; \quad |\delta\psi(t)\rangle = \sum_{n=1}^{\infty} \lambda(t)^n |\varphi_n\rangle \tag{5.12}$$

where $\xi_n$ and $|\varphi_n\rangle$ are time independent. Use this to obtain,

$$i\sum_{n=1}^{\infty} n\varepsilon\lambda(t)^n |\varphi_n\rangle = \sum_{n=1}^{\infty} \lambda(t)^n (H_0 - E_0)|\varphi_n\rangle + \lambda(t)QV|0\rangle + \sum_{n=2}^{\infty} \lambda(t)^n QV|\varphi_{n-1}\rangle - \sum_{n=2}^{\infty} \lambda(t)^n \sum_{m=1}^{n-1} \xi_{n-m}|\varphi_m\rangle \tag{5.13}$$

and,

$$\sum_{n=1}^{\infty} \lambda(t)^n \xi_n = \lambda(t)\langle 0|V|0\rangle + \sum_{n=2}^{\infty} \lambda(t)^n \langle 0|V|\varphi_{n-1}\rangle \tag{5.14}$$

From this we obtain,

$$|\varphi_1\rangle = \frac{-1}{(H_0 - E_0) - i\varepsilon} QV|0\rangle, \quad |\varphi_n\rangle = \frac{-1}{(H_0 - E_0) - in\varepsilon}\left[ QV|\varphi_{n-1}\rangle - \sum_{m=1}^{n-1} \xi_{n-m}|\varphi_m\rangle \right]; n = 2, 3, \ldots \tag{5.15}$$

where,

$$\xi_1 = \langle 0|V|0\rangle; \quad \xi_n = \langle 0|V|\varphi_{n-1}\rangle \text{ for } n = 2, 3, \ldots \tag{5.16}$$

The first few terms of these expansions are given in Appendix 3.

In the limit $\varepsilon \to 0$ the $|\varphi_n\rangle$ and $\xi_n$ terms are well behaved. For example,

$$|\varphi_2\rangle = \sum_{m \neq 0, n \neq 0} \frac{|n\rangle V_{nm} V_{m0}}{[(E_n - E_0) - 2i\varepsilon][(E_m - E_0) - i\varepsilon]} - \sum_{n \neq 0} \frac{|n\rangle V_{n0} V_{00}}{[(E_n - E_0) - 2i\varepsilon][(E_n - E_0) - i\varepsilon]} \quad (5.17)$$

In the limit $\varepsilon \to 0$ this becomes,

$$|\varphi_2\rangle \underset{\varepsilon \to 0}{=} \sum_{m \neq 0, n \neq 0} \frac{|n\rangle V_{nm} V_{m0}}{[(E_n - E_0)][(E_m - E_0)]} - \sum_{n \neq 0} \frac{|n\rangle V_{n0} V_{00}}{(E_n - E_0)^2} \quad (5.18)$$

We can see here why it is important that the initial state $|0\rangle$ be non-degenerate. If this was not the case then some of the terms in the denominator could be zero. However this will not occur in this case because we have specified that $E_n \neq E_0$ for $n \neq 0$.

A $(1/\varepsilon)$ divergence will appear in the phase when $G(t)$ is calculated. Use (5.8) along with (5.12) to obtain,

$$G(t) = \frac{1}{\varepsilon} \sum_{n=1}^{\infty} \frac{x^n e^{n\varepsilon t}}{n} \xi_n \quad (5.19)$$

Expand $\xi_n(\varepsilon)$ as a Taylor series,

$$\xi_n(\varepsilon) = \xi_n(0) + \frac{d\xi_n(0)}{d\varepsilon}\varepsilon + O(\varepsilon^2) \quad (5.20)$$

Therefore in the limit $\varepsilon \to 0$,

$$G(t) \underset{\varepsilon \to 0}{=} \frac{1}{\varepsilon} \sum_{n=1}^{\infty} \frac{x^n (1 + n\varepsilon t)}{n} \left( \xi_n(0) + \frac{d\xi_n(0)}{d\varepsilon}\varepsilon + O(\varepsilon^2) \right) \underset{\varepsilon \to 0}{=} \frac{G_a}{\varepsilon} + \Delta E t + iG_b \quad (5.21)$$

where,

$$G_a = \sum_{n=1}^{\infty} \frac{x^n \xi_n(0)}{n}, \quad \Delta E = \sum_{n=1}^{\infty} x^n \xi_n(0), \quad G_b = -i \sum_{n=1}^{\infty} \frac{x^n}{n} \frac{d\xi_n(0)}{d\varepsilon} \quad (5.22)$$

Use this in (5.8) to obtain,

$$|\psi(t)\rangle \underset{\varepsilon \to 0}{=} e^{-i(G_a/\varepsilon)} e^{-iE_0 t} e^{-i\Delta E t} e^{G_b} \left( |0\rangle + \sum_{n=1}^{\infty} x^n |\varphi_n\rangle_{\varepsilon=0} \right) \quad (5.23)$$

where $|\varphi_n\rangle_{\varepsilon=0}$ means that $|\varphi_n\rangle$ is evaluated at $\varepsilon = 0$. In the above $G_a$, $\Delta E$, and $G_b$ are all real and are non-divergent. Therefore the $(1/\varepsilon)$ terms are in a time-independent phase factor $\exp(-i(G_a/\varepsilon))$.

## 6. Conclusion.

We have examined a 2 state model system that was previously analyzed in Ref. [4]. We have shown that it is possible to obtain a well-defined solution for this system in the adiabatic limit where all the divergent terms appear in phase factor and that the result, at $t = 0$, is an eigenfunction of the Hamiltonian $H(0)$. This confirms that the adiabatic theorem holds for this case. We have also shown that in the general case, with an infinite number of states, the divergence that appear in the limit of adiabatic switching will also be confined to a phase factor.

## Appendix 1.
Some solutions to Eq. (4.7),

$$\tilde{g}_1(\varepsilon) = \frac{-i}{(2i\delta + \varepsilon)}, \quad \tilde{g}_2(\varepsilon) = -\frac{i}{(2i\delta + 3\varepsilon)(2i\delta + \varepsilon)^2}, \quad \tilde{g}_3(\varepsilon) = -\frac{2i}{(2i\delta + 5\varepsilon)(2i\delta + 3\varepsilon)(2i\delta + \varepsilon)^3},$$

$$\tilde{g}_4(\varepsilon) = -\frac{i}{(2i\delta + 7\varepsilon)}\left[\frac{1}{(2i\delta + 3\varepsilon)^2 (2i\delta + \varepsilon)^4} + \frac{4}{(2i\delta + 5\varepsilon)(2i\delta + 3\varepsilon)(2i\delta + \varepsilon)^4}\right] \tag{7.1}$$

## Appendix 2.
We want to show that $e^{F_b(0,0)} = 1/\sqrt{(1+(\Delta E/x)^2)}$. Use (4.9) in (4.5) to obtain,

$$g(t,\varepsilon) = \sum_{n=1}^{\infty} x^{2n} e^{2n\varepsilon t} \left(\tilde{g}_n(0) + \varepsilon r_n(\delta) + O(\varepsilon^2)\right) \tag{8.1}$$

and,

$$g'(t,\varepsilon) = \varepsilon \sum_{n=1}^{\infty} 2nx^{2n} e^{2n\varepsilon t} \tilde{g}_n(0) + O(\varepsilon^2) \tag{8.2}$$

Refer to (4.3) to obtain at $t=0$.

$$g'(0,\varepsilon) - ig(0,\varepsilon)^2 + (2i\delta - \varepsilon)g(0,\varepsilon) = -ix^2 \tag{8.3}$$

From the relationships in Section 4 we have,

$$\Delta E(x) = \sum_{n=1}^{\infty} x^{2n} \tilde{g}_n(0) \text{ and } F_b(0,0) = -i\sum_{n=1}^{\infty} \frac{x^{2n}}{2n} r_n(\delta) \tag{8.4}$$

From this we can obtain,

$$x\frac{d\Delta E}{dx} = \sum_{n=1}^{\infty} 2nx^{2n} \tilde{g}_n(0) \text{ and } ix\frac{dF_b(0,0)}{dx} = \sum_{n=1}^{\infty} x^{2n} r_n(\delta) \tag{8.5}$$

Use this to obtain,

$$g'(0,\varepsilon) = \varepsilon x \frac{d\Delta E}{dx} + O(\varepsilon^2) \text{ and } g(0,\varepsilon) = \Delta E + i\varepsilon x \frac{dF_b(0,0)}{dx} + O(\varepsilon^2) \tag{8.6}$$

Use the above result in (8.3) and equate terms to the 0$^{th}$ and 1rst order in $\varepsilon$ to obtain,

$$-\Delta E^2 + 2\delta \Delta E + x^2 = 0 \text{ and } 2x\frac{dF_b(0,0)}{dx}(\delta - \Delta E) + \left(\Delta E - x\frac{d\Delta E}{dx}\right) = 0 \tag{8.7}$$

The solution to these two equations are $\Delta E = \delta - \sqrt{\delta^2 + x^2}$ and $e^{F_b(0,0)} = 1/\sqrt{(1+(\Delta E/x)^2)}$.

## Appendix 3.
We calculate the first few terms of (5.15) and (5.16). Use $\sum_n |n\rangle\langle n| = 1$ and $V_{nm} = \langle n|V|m\rangle$ to obtain,

$$|\varphi_1\rangle = \frac{-1}{(H_0 - E_0) - i\varepsilon} QV|0\rangle = -\sum_{n\neq 0} \frac{|n\rangle V_{n0}}{(E_n - E_0) - i\varepsilon} \tag{9.1}$$

$$|\varphi_2\rangle = \sum_{m\neq 0, n\neq 0} \frac{|n\rangle V_{nm} V_{m0}}{\left[(E_n - E_0) - 2i\varepsilon\right]\left[(E_m - E_0) - i\varepsilon\right]} - \sum_{n\neq 0} \frac{|n\rangle V_{n0} V_{00}}{\left[(E_n - E_0) - 2i\varepsilon\right]\left[(E_n - E_0) - i\varepsilon\right]} \quad (9.2)$$

and,

$$\xi_1 = V_{00}, \quad \xi_2 = \langle 0|V|\varphi_1\rangle = \sum_{n\neq 0} \frac{V_{0n} V_{n0}}{(E_n - E_0) - i\varepsilon} \quad (9.3)$$

$$\xi_3 = \langle 0|V|\varphi_2\rangle = \sum_{m\neq 0, n\neq 0} \frac{V_{0n} V_{nm} V_{m0}}{\left[(E_n - E_0) - 2i\varepsilon\right]\left[(E_m - E_0) - i\varepsilon\right]} - \sum_{n\neq 0} \frac{V_{0n} V_{n0} V_{00}}{\left[(E_n - E_0) - 2i\varepsilon\right]\left[(E_n - E_0) - i\varepsilon\right]}$$

(9.4)